\newcommand{\g}{{g}}
\newcommand{\gs}{{g^{*}}}

\newcommand{\JI}{{\epsilon^{i_1 \cdots i_{p-2}}\frac{\partial
X_I^{a_1}}{\partial \sigma_I^{i_1}}\cdots \frac{\partial
X_I^{a_{p-2}}}{\partial \sigma_I^{i_{p-2}}}}}
\newcommand{\JJ}{{\epsilon^{j_1 \cdots j_{p-2}}\frac{\partial
X_{Ja_1}}{\partial \sigma_J^{j_1}}\cdots \frac{\partial
X_{Ja_{p-2}}}{\partial \sigma_J^{i_{p-2}}}}}
\newcommand{\Gai}{{\Gamma_I}}
\newcommand{\Gaj}{{\Gamma_J}}
\newcommand{\cV}{{\cal V}}
\newcommand{\cM}{{\cal M}}
\newcommand{\cO}{{\cal O}}
\newcommand{\bR}{{\bf R}}
\newcommand{\kakko}[2]{{\hat[#1,#2\hat]}}

\documentclass[12pt]{article}
\begin{document}
\renewcommand{\thefootnote}{\fnsymbol{footnote}}
\font\csc=cmcsc10 scaled\magstep1
{\baselineskip=14pt
 \rightline{
 \vbox{\hbox{UT-911}
        \hbox{October 2000}
 }}}

\vfill
\begin{center}
{\LARGE
Volume Preserving Diffeomorphism}\\
{\LARGE
and Noncommutative Branes
}

\vfill

{\csc Y. Matsuo}\footnote{
      e-mail address : matsuo@hep-th.phys.s.u-tokyo.ac.jp} and
{\csc Y. Shibusa}\footnote{
      e-mail address : shibusa@hep-th.phys.s.u-tokyo.ac.jp}
\vskip.1in

{\baselineskip=15pt
\vskip.1in
  Department of Physics,
  University of Tokyo\\
  Bunkyo-ku, Hongo 7-3-1\\
 Tokyo 113-0033\\
 Japan
\vskip.1in
}

\end{center}
\vfill

\begin{abstract}
{
We give a representation of
the volume preserving diffeomorphism
of $\bR^p$ in terms of the noncommutative $(p-2)$-branes
whose kinetic term is described by the Hopf term.
In the static gauge, the $(p-2)$-brane can be
described by the free fields and it suggests that
the quantization of the algebra
is possible.
}
\end{abstract}
\vfill

hep-th/0010040
\setcounter{footnote}{0}
\renewcommand{\thefootnote}{\arabic{footnote}}
\newpage

\section{Introduction}
In the context of the brane dynamics in the
string theory, the volume preserving diffeomorphism
of $p$-dimensions ($\cV_p$), infinitesimally generated by
\begin{equation}\label{vpd}
 \delta_v x^a = v^a(x),\quad
 \partial_a v^a(x)=0\,\,,
\end{equation}
is the important symmetry which reflects
the non-commutativity of the space-time.
The most famous example is the case $p=2$ when
the area preserving diffeomorphism
($\cV_2$) is the dynamical degree of
freedom in the matrix models \cite{r:BFSS}\cite{r:IKKT}.
In the BFSS matrix model \cite{r:BFSS}, such
a degree of freedom appears as discretization of
the residual gauge symmetry of the supermembrane
\cite{r:dWHN}. In the passage from \cite{r:dWHN}
to \cite{r:BFSS}, the quantized algebra of the symmetry
was essential to have the D0-brane interpretation.
It also appeared in the context
of the physics in large B limit and it becomes
one of the most actively studied branches in the string
theory \cite{r:CDS}\cite{r:SW}.

Intuitively the generalization to $p>2$ becomes important
when the antisymmetric $p$-form field is very large.  
Such a field is coupled to the volume of the $(p-1)$-branes.
When it is large, the configuration space is restricted to
the motion which preserves the volume element. In a related
but slightly different context, 
$\cV_p$ also appears as the  residual symmetry after the 
light-cone gauge fixing \cite{r:BSTT}.

Normally these symmetries have been studied from
the viewpoint the Nambu bracket
\cite{r:Nambu} which generalizes the Poisson bracket. 
This forces us to extend the idea of the phase space and requires
a radical change for the quantization.
{}For the recent mathematical studies, 
see for example \cite{r:T1}.
In such works, however, it is not clear to understand
the relation with the conventional quantization procedure.

On the other hand, in the discussion of the noncommutativity of 
$\cM$-theory, the $B$ field is replaced by 3rd order
antisymmetric tensor $C$ \cite{r:Li}. It is indicated in 
\cite{r:BBSS}\cite{r:KS} that as a natural extension of 
the usual noncommutative geometry,
we meet the extended object ``noncommutative string'' 
which appears at the end of open membrane.

In this paper, therefore, we use a different route
to define the quantization of $\cV_p$. Instead of
redefining the idea of the symplectic structure itself,
we use the extended objects ($(p-2)$-branes) to describe
the symmetry while we use the conventional phase space
and the quantization. Unlike the ordinary $D$-branes whose is 
described by the Dirac-Born-Infeld action, we use the action
which is similar to the Hopf term in the nonlinear
sigma model. Actually such an object appeared long before
in the context of the vortex motion in the fluid dynamics
for $p=3$ \cite{r:LR}.  

This paper is organized as follows. In section 2 we summarize 
the phase structure of 
the volume preserving diffeomorphism of $p$-dimensions ($\cV_p$).
In section 3 we give a description of the noncommutative
$(p-2)$-branes and show that they are reduced to a free
field theory in the static gauge.  We will demonstrate then that
a representation of $\cV_p$ is given in terms of the
noncommutative branes in this gauge. 
In the appendix \ref{vortex} we review the derivation of 
the action for $(p-2)$-branes as the action for the ``vortex'' 
of the fluid dynamics. In the appendix B we give a proof that 
$\cV_p$ thus derived is independent of the particular gauge fixing.

\section{Symplectic structure of Volume Preserving Diffeomorphism}
We introduce the symplectic structure  of
$\cV_p$ by the coadjoint orbit method \cite{r:Arnold}. 
The Lie algebra of $\cV_p$ (which we denote $\g$) 
is defined by  the commutation relation between two vector fields 
in (\ref{vpd}),
\begin{equation}\label{e_VPD0}
 [v,w]^a=v^b\partial_b w^a-w^b\partial_b u^a\,\,,
\end{equation}
where and in the following $a,b$ runs from 1 to $p$.
We use $U_a(x)$ 
which satisfies $\partial_a U^a=0$ as representing
the element in $\gs$ with the inner product,
\begin{equation}
 \langle U,v\rangle \equiv U_v \equiv \int d^p x U_a(x) v^a(x).
\end{equation}
The symplectic structure on the coadjoint orbit implies that
we have the following Poisson bracket between $U$,
\begin{equation}\label{e_VPD}
 \left\{U_a(x), U_b(y)\right\}=(\partial_b U_a(x)-
\partial_a U_b(x))\delta^{(p)}(x-y),
\end{equation}
which can be rewritten in terms of $U_v$ as,
\begin{equation}
 \left\{U_v,U_w\right\}=U_{[v,w]}.
\end{equation}
The last equation explicitly shows that (\ref{e_VPD}) gives
the representation of the algebra (\ref{e_VPD0}).

In the dynamics of the perfect fluid, this Poisson bracket
gives the Hamiltonian formalism of the
Navier-Stokes equation. Indeed if we choose
$$
 H=\frac{1}{2}\int d^p x U_a(x) U^a(x),
$$
as the Hamiltonian. Then the equation of motion
for $U$ is given by the Navier-Stokes equation,
\begin{equation}\label{Navier-Stokes}
 \frac{\partial U^a}{\partial t}=
\left\{H,U^a\right\}=U^b \partial^a U_b
 -U^b\partial_b U^a\,\,.
\end{equation}
In this sense, the Navier-Stokes equation
describes a straight line in $\cV_p$
and the Poisson bracket (\ref{e_VPD})
describes the symplectic structure
of the velocity fields.

As we review in the appendix, the noncommutative
brane describes the origin of the codimension
two vortex in $\bR^p$.
In this context, it is more convenient to rewrite 
the Poisson bracket in terms of the vorticity $(p-2)$ forms,
\begin{equation}
 \omega^{a_1 \cdots a_{p-2}}(x) \equiv \frac{1}{(p-2)!} 
\epsilon^{a_1 \cdots a_p} 
\partial_{a_{p-1}}U_{a_p}.
\end{equation}
We write the generators of $\cV_p$ in terms of $\omega$ as,
\begin{equation}
 \omega_f \equiv \int d^p x \epsilon_{a_1 \cdots a_p} \omega^{a_1 \cdots a_{p-2}}(x)
f^{a_{p-1}a_p}(x).
\end{equation}
$\omega_f$ is related to $U_v$ as,
\begin{equation}
 \omega_f = U_v,\quad \mbox{with}\quad
 v^a(x) = -2\partial_b f^{ba}.
\end{equation}
It is clear that the vector field $v^a$ thus defined from $f$
automatically satisfies $\partial_a v^a=0$.
There are, however,  some arbitrariness to define $f$ from $v$.
To fix it, we require that the two form $f_{ab}dx^a dx^b$
satisfies the equation $df=0$.  Together with the defining
equation $-2\delta f=v$ ($\delta=*d*$ with $*$ as Hodge dual), 
one can determine $f$ from $v$ as (up to the harmonic two forms),
\begin{equation}\label{Green}
 f_{ab}(x) = \frac{1}{2}\int d^p y \, G^{(p)}(x,y) \left(
\frac{\partial}{\partial y^a} v_b(y)
-\frac{\partial}{\partial y^b} v_a(y)
\right),
\end{equation}
with the Green function
$\partial^2_x G^{(p)}(x,y) =- \delta^{(p)}(x-y)$.
In terms of $\omega_f$, the Poisson bracket (\ref{e_VPD}) is replaced
by,
\begin{eqnarray}
 \left\{\omega_f,\omega_g\right\}      &  =     & 
\omega_{\kakko{f}{g}},\nonumber\\
 \left(\kakko{f}{g}\right)^{ab}   & \equiv & 
2\partial_c f^{cb}\partial_d g^{da}
-2\partial_c f^{ca}\partial_d g^{db}.
\label{e_VPD2}
\end{eqnarray}
It is easy to check the consistency condition between
two commutators, $-2\partial_b \left(\kakko{f}{g}\right)^{ba}=([v,w])^a$
if $-2\partial_b f^{ba}=v^a$ and 
$-2\partial_b g^{ba}=w^a$.
Although (\ref{e_VPD}) directly defines the volume
preserving diffeomorphism, we will mainly use (\ref{e_VPD2})
since the vorticity has much more direct interpretation
in terms of the non-commutative branes.

When $p=2$, (\ref{e_VPD2}) has a more familiar form.
We rewrite $f^{ab}(x)=\epsilon_{ab}f$ and so on,
the commutator (\ref{e_VPD2}) reduces to,
\begin{equation}
 \left(\kakko{f}{g}\right)^{ab} =\left\{f,g\right\}\epsilon^{ab},
  \quad
  \left\{f,g\right\} =\epsilon^{ab}\partial_a f \partial_b g,
\end{equation}
which is the conventional Poisson bracket for the coordinates
which produces the area-preserving diffeomorphism.

\section{Theory of Noncommutative Branes}
The starting point of our discussion 
for the noncommutative $(p-2)$-branes embedded in ${{\bf R}}^p$
is the action of the following form,
\begin{eqnarray}
\label{kinetic}
  S_0  = \frac{-1}{p} \sum_{I=1}^N \Gamma_I
           \int d\sigma_I dt\,\epsilon_{a_1 \cdots a_p}
            X_I^{a_1} \frac{\partial X_I^{a_2}}{\partial t} 
           \epsilon^{i_1 \cdots i_{p-2}}
            \frac{\partial X_I^{a_3}}{\partial \sigma_I^{i_1}}\cdots
            \frac{\partial X_I^{a_p}}{\partial \sigma_I^{i_{p-2}}}\,\,,
\end{eqnarray}
where $\sigma^i_I$ are the coordinates for the spatial
direction of $I$-th $(p-2)$-brane. Here we wrote
only the kinetic term which governs the phase space.
One may add potential terms which do not contain
the time derivative of $X$ to (\ref{kinetic})
without changing the algebra.

We first recall how such an action appeared
in the context of string theory.
For $p=2$, 
it describes the motion of the end points of the open string
on the D-brane \cite{r:CH},
\begin{equation}
 S=\frac{1}{2} \int_\Sigma d^2\sigma 
B_{ab}\epsilon_{\alpha\beta}
\partial_\alpha X^a \partial_\beta X^b
  = \frac{1}{2} \int_{\partial \Sigma} dt 
B_{ab} X^a \partial_t X^b.
\end{equation}
If $B$ field is block diagonalized,
$\Gamma_I$ is identified with $I$-th block of \newline
$B=diag\left(
\Gamma_1 \epsilon,\cdots\Gamma_N\epsilon\right)$ with $\epsilon=
\left(
\begin{array}{c c}
 0&1 \\ -1 & 0
\end{array}\right)$.
Similarly, for $p=3$, such a kinetic term appeared
in the description 
of ``noncommutative string'' \cite{r:BBSS}\cite{r:KS}
which describes the stringy degree of freedom appearing
at the boundary of the open membrane world volume.
In this case $\Gamma_I$ is related to the magnitude of the 
3rd order anti-symmetric tensor in $\cM$-theory.
The $(p-2)$-branes which is described by the action
(\ref{kinetic}) is a generalization of these
extended noncommutative objects. In this context 
we call them as ``noncommutative $p$-brane''.

The noncommutative $p$-brane for $p=0,1$ actually 
appeared long before in the literature of
the fluid dynamics.  As we mentioned in the
previous section, the volume preserving diffeomorphism
is the configuration space for the perfect fluid and
Euler's equation describes the straight line in $\cV_p$
\cite{r:Arnold}.  The noncommutative brane
of codimension two describes the vortex which keeps its form
under time evolution, and therefore it is possible
to derive the action which describes its own motion.  
See the appendix \ref{vortex} for the derivation of the
kinetic term (\ref{kinetic}).  For $p=3$, such an action
was discussed in the classic paper \cite{r:LR}. For the geometrical
and dynamical implication of the Hopf term, see for example 
\cite{r:Pol}\cite{r:Hopf}.

Let us proceed to the analysis of the symplectic
structure defined by $S_0$.
Since $S_0$ contains only one time derivative,
the momentum variable is written in terms of $X$,
\begin{equation}
 \Pi_{Ia}(\sigma_I)\equiv \frac{\partial S_0}{\partial \dot{X_I^a}}(\sigma_I)=
  \frac{\Gai}{p}\epsilon_{ab a_1\cdots a_{p-2}}
  X_I^b\JI (\sigma_I).
\end{equation}
Since the conjugate variable is written by the original fields,
we need to impose the primary constraints,
\begin{equation}\label{primary}
 \phi_{Ia}(\sigma_I) =\Pi_{Ia}(\sigma_I)-
  \frac{\Gai}{p}\epsilon_{ab a_1\cdots a_{p-2}}
  X_I^b\JI (\sigma_I)\approx 0.
\end{equation}
By using the canonical commutation relation,
\begin{equation}
 \left\{X_I^a(\sigma_I),\Pi_{Jb}(\sigma_J')\right\}=\delta_{IJ}\delta^a_b\delta^{(p-2)}
 (\sigma_I-\sigma_J')\,\,,
\end{equation}
the Poisson bracket between $\phi_{Ia}$ is given by,
\begin{eqnarray}
 \left\{\phi_{Ia}(\sigma_I),
\phi_{Jb}(\sigma'_J)\right\} &=     & 
-\Gai \delta_{IJ}\epsilon_{aba_1
        \cdots a_{p-2}}
        \JI \nonumber \\
&      &\cdot\,\delta^{(p-2)}(\sigma_I-
       \sigma'_I) \nonumber\\
            &\equiv& M^I_{ab}\delta_{IJ}\delta^{(p-2)}
            (\sigma_I-\sigma'_I)\,\,.
\end{eqnarray}
Since $M^I_{ab}\frac{\partial X^b_I}{\partial \sigma_I^i}=0$
for $i=1,2,\cdots,p-2$, the matrix $M$ has rank 2.
We therefore expect to have $p-2$ first class constraints
which correspond to the reparametrization of the
world volume coordinates $\sigma$.

Indeed if we combine them into
$T_{Ii}=-\frac{\partial X_I^a}{\partial \sigma_I^i}\phi_{Ia}$, they
satisfy the algebra,
\begin{eqnarray}
  \left\{T_{Ii}(\sigma_I),T_{Jj}(\sigma'_J)\right\}&=&
   \delta_{IJ}T_{Ii}(\sigma_I)\frac{\partial}{\partial \sigma_I^j}
   \delta^{(p-2)}(\sigma_I-\sigma'_J)\nonumber\\
  &&  + \delta_{IJ}
   T_{Ij}(\sigma_I)\frac{\partial}{\partial \sigma_I^i}
   \delta^{(p-2)}(\sigma_I-\sigma'_J)\nonumber \\
  && +\delta_{IJ}\frac{\partial T_{Ii}(\sigma_I)}{\partial \sigma_I^j}
   \delta^{(p-2)}(\sigma_I-\sigma'_J),
\end{eqnarray}
which is the generalization of the 
classical Virasoro algebra to the higher dimensions.
The existence of the first class constraints 
forces us to introduce some extra constraints to fix the gauge.


There are several possible choices for the gauge fixing.
The choice we took in this paper
is the {\em static gauge}\footnote{
For $O(3)$ invariant but nonlinear constraints for $p=3$,
see \cite{r:BBSS}\cite{r:KS}\cite{r:Mat}.
} defined by,
\begin{equation}\label{constraint2}
 \chi_I^i\equiv X_I^{i+2}(\sigma_I) -\sigma_I^i\approx 0\,\,,
\end{equation}
where (and in the following) $i$ takes its value
in the range $1,\cdots,p-2$.
It was shown that the set of constraints $\left\{\xi_\alpha\right\}
=\left\{
\phi_{Ia}(\sigma_I),\chi_J^i(\sigma_J)\right\}$,
($\alpha$ represents the indices $I,i,a,\sigma_I$) becomes the second class
and we can define the Dirac bracket as,
\begin{equation}
 \left[F,G\right]_D = \left\{F,G\right\}
 -\sum_{\alpha,\beta} \left\{F,\xi_\alpha\right\}
 (C^{-1})^{\alpha\beta}\left\{\xi_\beta,G\right\},
\end{equation}
where the matrix $C$ is defined as $C_{\alpha\beta}=\left\{
\xi_\alpha, \xi_\beta\right\}$.
In this gauge choice, we obtain 
\begin{eqnarray}
\label{cin}
 C^{-1}[\chi_{Ii}
 (\sigma_I),\phi^a_J
 (\tilde{\sigma}_J)] &=& -\delta_{IJ}\frac{\partial X_I^a(\tilde{\sigma}_I)}
                         {\partial \tilde{\sigma}_I^i}
                         \Theta(\tilde{\sigma}_I^i-\sigma_I^i)
                         \prod_{j=1}^{i-1}\prod_{j=i+1}^{p-2}
                         \delta(\sigma_I^j-\tilde{\sigma}_I^j), \nonumber \\
 C^{-1}[\phi^a_I
 (\sigma_I),\phi^b_J
 (\tilde{\sigma}_J)] &=& \frac{1}{\Gai (p-2)!}\epsilon^{ab34\cdots p}\delta_{IJ}
                         \delta^{(p-2)}(\sigma_I-\tilde{\sigma}_J), \nonumber \\
  C^{-1}[\chi_{Ii}
 (\sigma_I),\chi_
 {Jj}
 (\tilde{\sigma}_J)] &=& 0.
\end{eqnarray}
It gives the bracket of {\em the free field theory},
\begin{eqnarray}
\label{daiji}
 \left[X_I^1(\sigma_I),X_J^2(\sigma'_J)\right]_D&=&
\frac{1}{\Gai (p-2)!}\delta^{(p-2)}(\sigma_I-\sigma'_J)\delta_{IJ},\nonumber\\
  \left[X_I^{i+2}(\sigma_I),X_J^a(\sigma'_J)\right]_D&=&0\,\,.
\end{eqnarray}
Actually the reduction to the free theory can be easily seen
at the Lagrangian level by putting the constraint
(\ref{constraint2}) to the action (\ref{kinetic}).
By just inspection, one can understand that it reduces
to the quadratic in terms of $X$ and thus define
a free theory. It reminds us of the fact that 
the ordinary string theory was 
defined by the nonlinear Nambu-Goto action but
it reduces to the free theory in the light-cone gauge.


As we explain the appendix, the vorticity is related to
the embedding functions of the $(p-2)$-brane as,
\begin{eqnarray}
\label{generators}
 \omega^{a_1 \cdots a_{p-2}}(x) \equiv 
 \sum_{I=1}^N \Gamma_I \int d^{p-2}\sigma_I \epsilon^{i_1\cdots i_{p-2}}
 \frac{\partial X^{a_1}_I}{\partial \sigma_I^{i_1}}\cdots
 \frac{\partial X^{a_{p-2}}_I}{\partial \sigma_I^{i_{p-2}}}
\delta^{(p)}(x-X_I(\sigma_I))
\,\,.
\end{eqnarray}
We identify the generator of the volume preserving diffeomorphism of 
$p$-dimensions ($\cV_p$) as follows,
\begin{eqnarray}
 \omega_f &\equiv& \int d^p x \epsilon_{a_1 \cdots a_p} \omega^{a_1 \cdots a_{p-2}}
(x)f^{a_{p-1}a_p}(x), \nonumber \\
 &=& \sum_{I=1}^{N}\Gai \int d\sigma_I\epsilon_{a_! \cdots a_{p-2}}\JI f^{a_{p-1}a_p}
    (X_I(\sigma_I)).
\end{eqnarray}
Indeed it generates $\cV_p$  on $X$,
\begin{eqnarray}
 \left[X_I^a(\sigma_I) , \omega_f \right]_D &=& 
-2\partial_bf^{ba}(X_I(\sigma_I))
 +\sum_{i=1}^{p-2}2\partial_bf^{b \,i+2}(X_I(\sigma_I))\frac{\partial X_I^a}
 {\partial \sigma_I^i}\label{volpre}\\
&=& v^a(X_I(\sigma_I)) + \left.
\frac{d}{dt}X^a(\sigma_I^i-t v^{i+2}(X_I(\sigma_I)))\right|_{t=0}\,\,.
\label{volpre2}
\end{eqnarray}
The second term in the right hand side of (\ref{volpre})
comes from the reparametrization
because the gauge choice (\ref{constraint2}) must be preserved. 
In the second line, we show that it can be absorbed in
the infinitesimal reparametrization of the world volume coordinates.

{}From (\ref{daiji}), one can show by the direct calculation 
that $\omega_f$ satisfies
the algebra of $\cV_p$,
\begin{equation}
\label{daiji2}
  \left[\omega_f , \omega_g \right]_D =- \omega _\kakko{f}{g}\,\,.
\end{equation}
where $\left(\kakko{f}{g}\right)^{ab}$ is defined in (\ref{e_VPD2}).
This algebra is consistent with $\cV_p$ because in this system
\begin{eqnarray}
 \left(\delta_f \delta_g - 
 \delta_g \delta_f \right) X^a &=& \left[\left[X^a ,\omega_g\right]_D , \omega_f\right]_D -
                                   \left[\left[X^a ,\omega_f\right]_D , \omega_g\right]_D \nonumber \\
                               &=& \left[X^a , \left[\omega_g , \omega_f\right]_D\right]_D \,\,.
\end{eqnarray}
Since $\omega_f$ is invariant under the
reparametrization of $\sigma$, we expect that
this result does not depend on the gauge choice.
We give some further reasoning for the gauge independence 
in appendix B.

At this point, it seems attractive if one may have a mapping
from arbitrary functions on $\bR^p$ to the functional 
of the brane coordinates such that we have the following
relation,
\begin{equation}
 F(x)\in C(\bR^p)  \rightarrow  \cO_F,\quad \mbox{such that}
\quad  \left\{\cO_F,\omega_f\right\}_D  =  \cO_{\delta_f F}\,\,.
\end{equation}
A plausible candidate is,
$\cO_F = \int d^{p-2}\sigma F(X(\sigma))$.
This is not, unfortunately, so straightforward.
As we explicitly wrote in (\ref{volpre2}), there are
extra term in the Dirac bracket which can be only preserved
in the reparametrization of $\sigma$. The expression for
$\cO_F$ can transform covariantly only when $F(X(\sigma))$
transforms as 
$d^{p-2}\sigma F(X(\sigma))=d^{p-2}\tilde\sigma F(X(\tilde\sigma))$.
Such a transformation is possible only if $F$ is $(p-2)$- (or 
Hodge dual $2$-)
form in the target space. In this case, they are
 actually the generators of $\cV_p$ themselves.
To define the covariant functional of the general $q$-forms,
we need to introduce the world volume metric tensor.

\section{Discussions}
In this paper, we gave the
classical symplectic structure of the volume preserving
diffeomorphism while not explicitly attempted to quantize
the symmetry.  The symplectic structure
for the noncommutative branes turned out to be very simple 
in the static gauge and
can be quantized immediately.  For $p=3$ case as
an example, the quantized version of the Dirac bracket 
(\ref{daiji}) gives the commutation relation
\begin{equation}
 \left[X^1(\sigma),X^2(\sigma')\right]
 = \delta(\sigma-\sigma').
\end{equation}
This commutation relation is the same as the
bosonic ghost in the superstring theory.
The generator of the volume preserving diffeomorphism
takes the following form,
\begin{equation}
 \omega_f = \int d\sigma \left(
f^{12}(X)+ f^{23}(X)\frac{\partial X^1}{\partial \sigma}
+f^{31}(X)\frac{\partial X^2}{\partial \sigma}
\right)\,\,.
\end{equation}
Somewhat similar
construction of space-time reparametrization symmetry
appeared in \cite{r:GKS} where generators are written
in terms of $\beta-\gamma$ system.
Unlike that situation, the generators of $\cV_3$ are nonlinear
both in $\beta$ and $\gamma$ and highly singular.
One obvious difficulty
is how to organize the regularization scheme 
such that we can keep the desirable symmetry  such as $O(3)$. 
The other, but maybe related, issue for $p=3$ case is the BRST invariance.
These considerations may give the constraints
on the target space.  

Another issue is how to introduce
the target space supersymmetry.  
In \cite{r:BBSS} \cite{r:KS}, these noncommutative
strings are introduced to describe the noncritical
self-dual string in six dimensions \cite{r:LS}.
Supersymmetry will be indispensable to
give some insights to these issues from our viewpoint.

\vskip 5mm

\noindent{\em \small Acknowledgements:
One of the author (YM) would like to thank 
K. Fujikawa, S. Yahikozawa,
and D. Minic for the comments and discussions.  He would
also like to thank I. Bars for his hospitality during
the stay at the CIT-USC center where the final part of this
draft was typed.

YM is supported in part by Grant-in-Aid 
(\# 09640352) and in part by Grant-in-Aid for Scientific
Research in a Priority Area, ``Supersymmetry and Unified
Theory of Elementary Particle'' (\# 707) from
the Ministry of Education, Science, Sports and Culture.}

\vskip 5mm

\appendix
\section{Vortices in Ideal Fluid}\label{vortex}

In fluid dynamics without viscosity, it is known that the
vortex configuration of codimension 2 (at least for $p=2,3$),
\begin{equation}\label{e_brane}
 \omega^{a_1 \cdots a_{p-2}}(x) = 
  \sum_{I=1}^N \Gai \int d\sigma_I \JI \delta^{(p)}(x-X_I(\sigma_I)).
\end{equation}
keeps its form under the time evolution.
To show it, we derive the vector field which
produces the vorticity as,
\begin{eqnarray}
 U_a(x) &=&(-1)^p
           \sum_{I=1}^N \Gai
            \int d\sigma_I \epsilon_{a a_1\cdots a_{p-2}b} \nonumber \\
        & &\cdot \,\JI \partial^b G^{(p)}(x-X_I(\sigma_I)),
\end{eqnarray}
with the Green function defined after (\ref{Green}).
If we put these expression into the Navier-Stokes
equation (\ref{Navier-Stokes}), 
one may show that a consistent solution can be found 
by keeping the vortex configuration itself while
the location of the vortex is changed by the equation
\begin{equation}\label{vortex_eq}
 \epsilon_{a_1\cdots a_p}
 \frac{\partial X_I^{a_1}}{\partial \sigma_I^1}
  \cdots
  \frac{\partial X_I^{a_{p-2}}}{\partial \sigma_I^{p-2}}
\left(U^{a_{p-1}}(X_I)-
\frac{\partial X_I^{a_{p-1}}}{\partial t}\right)=0\,\,,
\end{equation}
or by solving it,
\begin{equation}\label{vortex_eq2}
 \frac{\partial X_I^a(t,\sigma_I)}{\partial t}
  = U^a(X_I(\sigma_I)) +\sum_{i=1}^{p-2} 
 \alpha^i_I\frac{\partial X^a_I(t,\sigma_I)}{\partial \sigma_I^i}.
\end{equation}
Here $\alpha^i_I$ is the arbitrary parameter. 
The second term represents
the time evolution along the brane and can be
absorbed into the time dependent reparametrization of world
brane coordinates $\sigma^{i}_I$. 

Originally the Navier-Stokes equation has the phase
space described by $U^a$, the above statement
shows that it can be consistently truncated to the much
smaller degree of freedom, namely the locations of $(p-2)$-branes.
It is then natural to suspect that there is the action
for $X$ which directly gives (\ref{vortex_eq}).
Such an equation has been known for $p=2,3$ \cite{r:LR},
and we give the higher dimensional extension here.
Indeed it is the combination of
the kinetic term (\ref{kinetic}) 
and the potential term which is specific of the
Navier-Stokes equation,
\begin{eqnarray}
  S    &=& S_0 - H \nonumber \\
  H    &=& \frac{1}{2} \sum_{I,J=1}^N \Gai \Gaj (p-2)! \int d\sigma_I d\sigma_J \JI \nonumber \\
       & &   \JJ G^{(p)}(X_I(\sigma_I)-X_J(\sigma_J)) \\
  S_0  &=& - \frac{1}{p} \sum_{I=1}^N \Gamma_I
           \int d\sigma_I dt\,\epsilon_{a_1 \cdots a_p}
            X_I^{a_1} \frac{\partial X_I^{a_2}}{\partial t} \epsilon^{i_1 \cdots i_{p-2}}
            \frac{\partial X_I^{a_3}}{\partial \sigma_I^{i_1}}\cdots
            \frac{\partial X_I^{a_p}}{\partial \sigma_I^{i_{p-2}}}\,\,.
\nonumber
\end{eqnarray}
The variation of this action gives (\ref{vortex_eq}).

Appearance of the undetermined parameters in
(\ref{vortex_eq2}) can be explained by solving
the constrained system.
The Hamiltonian for $(p-2)$-branes, $H$, should
be modified by including the constraints (total Hamiltonian),
\begin{eqnarray}
 H_T \equiv H + \sum_{I=1}^N\int d\sigma_I
\lambda_I^a(\sigma_I)\phi_{Ia}(\sigma_I).
\end{eqnarray}
{}From the consistency condition
$\dot{\phi_{Ia}}=\left\{\phi_{Ia},H_T \right\} \approx 0$, we obtain
\begin{eqnarray}
 \lambda_I^a(\sigma_I)=u^a(X_I(\sigma_I))+\sum_{i=1}^{p-2} \alpha^i(\sigma_I)
 \frac{\partial X^a_I(\sigma_I)}{\partial \sigma_I^i}.
\end{eqnarray}
Here $\alpha^i(\sigma_I)$ are undetermined functions. 
This ambiguity comes from the reparametrization invariance.
It is fixed by the gauge conditions (\ref{constraint2}), 
\begin{equation}
\alpha^i(\sigma_I) = -u^{i+2}(X_I(\sigma_I)) + a^i,
\end{equation}
where $a^i$ are constant parameters which correspond to $p$-brane translation.

\section{Gauge independence of $\cV_p$}
In this appendix, we give a confirmation that
the algebra (\ref{daiji2}) does not depend on
the gauge fixing condition (\ref{constraint2}).
It may be obvious that the Dirac bracket between
the gauge invariant quantities does not depend on
the particular gauge choice.  However, since
we do not know the explicit proof of this statement,
we give a (admittingly insufficient) calculation
which support it.

We change the gauge fixing condition infinitesimally by,
\begin{eqnarray}
\label{constraint3}
 \chi_I^i &\equiv& X_I^{i+2}(\sigma_I) -\sigma_I^i\approx 0,
\qquad (i=2,\cdots,p-2) \\
 \chi_I^3 &\equiv& X_I^3(\sigma _I) - \sigma_I^1 
-\epsilon h(X_I(\sigma_I),\sigma_I)\approx 0.
\end{eqnarray}
Up to the first order of $\epsilon$, the Dirac bracket
between $X$ is modified to
\begin{eqnarray}
\label{daiji3}
 \left[X_I^1(\sigma_I),X_J^2(\sigma'_J)\right]_D&=&
\frac{1}{\Gai (p-2)!}\delta^{(p-2)}(\sigma_I-\sigma'_J)\delta_{IJ} \nonumber\\
&& +\epsilon \frac{1}{\Gai (p-2)!}\delta^{(p-2)}(\sigma_I-\sigma'_J)\delta_{IJ}
   \left(-\frac{\partial h}{\partial \sigma_I^1}-\frac{\partial h}
   {\partial X_I^3}\right), \nonumber \\
  \left[X_I^1(\sigma_I),X_J^3(\sigma'_J)\right]_D&=&\epsilon \frac{1}{\Gai (p-2)!}
  \delta^{(p-2)}(\sigma_I-\sigma'_J)\delta_{IJ}
   \left(\frac{\partial h}
   {\partial X_I^2}\right), \nonumber \\
  \left[X_I^2(\sigma_I),X_J^3(\sigma'_J)\right]_D&=&\epsilon \frac{1}{\Gai (p-2)!}
  \delta^{(p-2)}(\sigma_I-\sigma'_J)\delta_{IJ}
   \left(-\frac{\partial h}
   {\partial X_I^1}\right), \nonumber \\
 \mbox{Others}&=&0\,\,.
\end{eqnarray} 
{}We confirmed that the
algebra (\ref{daiji2}) does not change to the first order
in $\epsilon$ with this modified brackets.


\end{document}